\documentclass[letters,usenatbib]{mnras}

\usepackage{newtxtext,newtxmath}

\usepackage[T1]{fontenc}
\usepackage{xcolor}
\DeclareRobustCommand{\VAN}[3]{#2}
\let\VANthebibliography\thebibliography
\def\thebibliography{\DeclareRobustCommand{\VAN}[3]{##3}\VANthebibliography}

\usepackage{graphicx}	
\usepackage{amsmath}	

\title[ \title{The time evolution of the ultraviolet habitable zone}]{The time evolution of the ultraviolet habitable zone}

\author[R. Spinelli et al.]{
R.~Spinelli,$^{1,2}$\thanks{E-mail: riccardo.spinelli@inaf.it} 
F.~Borsa,$^{3}$ 
G.~Ghirlanda,$^{3,4}$ 
G.~Ghisellini,$^{3}$ 
F.~Haardt,$^{2,3,4}$ 
and F.~Rigamonti$^{2,3,4}$ \\
$^{1}$INAF -- Osservatorio Astronomico di Palermo, Piazza del Parlamento 1, 90134 Palermo (PA), Italy\\
$^{2}$Dipartimento di Scienza e Alta Tecnologia, Universit\`a dell'Insubria, Via Valleggio 11, 22100 Como, Italy\\
$^{3}$INAF -- Osservatorio Astronomico di Brera, Via E. Bianchi 46, 23807 Merate (LC), Italy \\
$^{4}$INFN -- Sezione Milano--Bicocca, Piazza della Scienza~3, 20126 Milano, Italy \\
}

\date{Accepted 2024 July 2. Received 2024 July 2; in original form 2024 January 15
}

\pubyear{2024}

\begin{document}
\label{firstpage}
\pagerange{\pageref{firstpage}--\pageref{lastpage}}
\maketitle

\begin{abstract}
{For stars hosting Circumstellar Habitable Zone (CHZ) exoplanets, we investigate the time-evolution of their ultraviolet habitable zone (UHZ), the annular region around a star where an exoplanet could experience a suitable ultraviolet environment for the presence and emergence of life, and the possible intersection of the UHZ with the CHZ. To estimate their UV luminosity evolution, and therefore the evolution of their UHZ, we analyse Swift-UV/Optical telescope observations and adopt the near-UV luminosity evolutionary tracks derived using GALEX observations of young moving groups. We find that an intersection between CHZ and UHZ could exist (or have existed) around all stars of our sample at different epochs, except for the coldest M-dwarfs (temperature $\lesssim2800$\, K, e.g. Trappist-1). For hotter M-dwarfs the formation of RNA precursors through cyanosulfidic chemistry triggered by near-UV radiation could occur during the first $\simeq$1-2\, Gyrs. The radial-extension and time-duration of the CHZ-UHZ intersection increase with the stellar effective temperature and the exoplanet atmospheric transmissivity at near-UV wavelengths. Within our sample, Proxima Centauri represents a golden target for the quest of life outside the Solar system  because it experienced a long-lasting and more extended, compared to similar M-dwarfs, CHZ-UHZ intersection. }
\end{abstract}




\section{Introduction}

The quest for life outside the Solar system represents one of the primary goals of exoplanetary research. Due to our uncertain understanding of the actual conditions necessary for the emergence and the presence of life, the search for potentially habitable planets involves a delicate balance between the very definition of habitability and the limitations posed by current and next-generation observational facilities. 

Currently, planets around M--dwarfs may represent the most promising targets to find habitable worlds. M--dwarfs represent the most abundant stellar population in our Galaxy ($\sim$75\%, \citealt{Bochansky2010}) and their physical properties, as their small radii, low masses and low luminosities facilitate the discovery and characterization of habitable zone exoplanets \citep[][]{Kasting1993, Kopparapu2013} through methods based on radial velocities and transits \citep{Nutzman2008, Quirrenbach2014}. For this reason, the majority of rocky planets within the habitable zone discovered to date orbit around M-dwarfs. 

On the theoretical side, it is worth investigating whether life as-we-know-it can emerge and persist on such worlds. Due to its importance for life on Earth, liquid water is considered an essential factor. For this reason, a rocky planet is considered potentially habitable if it resides in the so-called Circumstellar Habitable Zone \citep[CHZ, e.g.,][]{Kasting1993, Kopparapu2013}, namely the annular region around a star in which there are suitable temperatures for the persistence of liquid water on the planetary surface. This region is primarily determined by the stellar properties (such as luminosity and effective temperature $T_{\rm eff}$) and the star-planet separation, but it can also depend on other factors such as planetary mass \citep{Kopparapu2014}, tidal locking \citep{Driscoll2015, Barnes2017}, atmospheric composition \citep{Goldblatt2013, Ramirez2014, Chaverot2023} and volcanic activity \citep{Ramirez2017}. In addition to the existence of liquid water, other intrinsic factors such as ongoing tectonics and volcanism may contribute to making a planet habitable (see \citealt{Meadows2018} for a review). 

Also, the high-energy radiation environment, determined by the host star and more rarely by high-energy transient events (\citealt{Ruderman1974,Thorsett1995,Gehrels2004,Melott2011, Piran2014, Spinelli2021}, see \citealt{SpinelliGhirlanda23} for a review), influences habitability. The impact of the high-energy radiation on habitability could be twofold. UV/X-ray radiation may cause atmospheric evaporation \citep{SanzForcada2010, Spinelli2023b, Maggio2023} and biomolecules destruction \citep{Sag, Buccino2007}. On the other hand,  experimental studies \citep[e.g.,][]{Toupance1977, Powner2009, Ritson2012, Patel2015, Xu2018, Rimmer2018} shows that UV light is crucial for the synthesis of some of the building blocks of life, up to and including ribonucleic acid (RNA).

In \citet{Spinelli2023} (hereafter S23) we studied and combined the positive (prebiotic) and negative (atmospheric evaporation and biomolecules destruction) roles of stellar UV radiation on potentially habitable exoplanets in order to estimate the Ultraviolet Habitable Zone (UHZ). This is the annular region around a star within which an exoplanet receives a UV flux suitable for the possible emergence and presence of life as-we-know-it on the planetary surface. In that work, by exploiting the existing public observations of the Ultra-Violet Optical Telescope (UVOT - \citealt{Roming2005}) on board the Neil Gehrels {\it Swift} Observatory \citep{Gehrels2004} of a sample of 17 stars harboring 23 planets in their CHZ, we found that the \emph{current} NUV luminosity of M–dwarf hosts in our sample is decisively too low to trigger abiogenesis – through cyanosulfidic chemistry \citep{Rimmer2018} – on their CHZ planet. Since our sample was mainly composed of old stars (age > 2.4 Gyrs), in S23 we had conjectured the possibility that during the early stages of their evolution, these M-dwarfs were brighter in the NUV band, irradiating their CHZ exoplanets with a flux able to trigger the formation of RNA precursors through cyanosulfidic chemistry. 

In this work, we test this hypothesis by exploiting the results obtained through GALEX observations of stellar young moving groups (\citealt{HAZMATIX} - R23 hereafter). We estimate the evolution of the CHZ and the UHZ of planets hosted by the stars of S23 in order to explore the existence and extent of a possible intersection of their CHZ and UHZ in the past. 

In Sect. \ref{sec:sample} we describe the planetary systems sample. The description of the methods adopted to estimate the time-evolution of the CHZ and the UHZ are provided in Sect. \ref{sec:chz}. In Sect. \ref{sec:discussion} we estimate the radial extension and the duration of intersection between CHZ and UHZ for the selected stars over time. Results are discussed in Sect. \ref{sec:discussion}.

\section{Sample selection}
\label{sec:sample}
We consider the stellar sample of S23. It contains planetary systems that host rocky planets in the CHZ and for which NUV observations performed by the Swift telescope are available. From this sample, we selected stars with an estimated age (or lower limit) reported in the literature and with stellar mass $<$0.9 $\rm M_{\odot}$. The properties of the selected 14 planetary systems are reported in Table \ref{tab:sample}. 

As in S23, we adopt the optimistic limits of the CHZ, empirically determined by \citet{Kasting1993} and \citet{Kopparapu2013}. These limits are based on the inferred presence of liquid water on Mars' surface before 3.8 Gyr ago \citep{Pollack1987,Bibring2006}, and on the absence of liquid water on Venus' surface for, at least, the past Gyr \citep{Solomon1991}. It is assumed (optimistically) that Venus and Mars actually had liquid water on their surfaces, respectively 1 and 3.8 Gyr ago and, therefore, that they were habitable at those epochs.

\section{CHZ and UHZ evolution}
\label{sec:chz}
To evolve the CHZ during the stellar lifetime we use the  MESA Isochrones and Stellar Tracks (MIST, \citealt{MIST0, MIST1}) for each stellar mass. Each evolutionary track provides several properties of the star at each time step from the pre-main sequence phase to the white dwarf cooling phase. We use the bolometric luminosity and the effective temperature to estimate, as a function of time, the values of the inner and the outer boundaries of the CHZ through the method described in \citet{Kopparapu2014}.

To estimate the evolution in time the UHZ defined in S23 we used the results obtained by R23 regarding the high-energy evolution of M and K stars. This work is based on GALEX observations of M and K star members of young moving groups (YMGs) and clusters ranging in age from 10 Myr to 650 Myr. Young moving groups are loose stellar associations for which ages can be reliably assessed using stellar isochrones and lithium depletion models. To this sample, R23 added M and K stars within 30 pc not identified as members of any known moving group or close-in binaries. To these latter stars, R23 assigned an age of 5 Gyr (the average age of the field stars nearby to our Sun). 

R23 reported the NUV fluxes observed with GALEX at an absolute distance of 10 pc of this sample and calculated the median of the NUV fluxes at each age bin (16 Myr, 43 Myr, 150 Myr, 650 Myr, 5 Gyr). 
They combined the 10 Myr and 24 Myr samples into a single age bin due to a low number of data points at the early ages. They divided the stars in this sample into three samples: late-M-dwarfs (0.1-0.35 $M_{\odot}$), early-M-dwarfs (0.35-0.6 $M_{\odot}$) and K stars (0.6-0.9 $M_{\odot}$).

Since the NUV band considered by \citet{Rimmer2018} is 200-280 nm, we used the following method to rescale the GALEX (177-283 nm) and the Swift (170-290 nm) observed flux. We convert the GALEX flux density in Jansky to the flux density in erg cm$^{-2}$ s$^{-1}$ A$^{-1}$ using the mean effective NUV wavelengths derived by \citet{Schneider2018} for M--dwarfs (255 nm) and \citet{HAZMATIX} for K stars (262 nm) that take into account the spectral shape of M and K stars in the total GALEX bandwidth (177-283 nm). Then we assume a flat spectrum (in erg cm$^{-2}$ s$^{-1}$ A$^{-1}$) in the NUV band (200-280 nm), as supported by visual inspection of archival NUV data obtained by HST/STIS and spectra delivered by the treasury MUSCLES survey \citep{2016ApJ...820...89F} and we estimate the integrated flux in the 200-280 nm by multiplying the obtained flux density by 80 nm. For Swift observations, we convert the count rate to flux density in erg cm$^{-2}$ s$^{-1}$ A$^{-1}$ using the conversion factor derived by \citet{Brown2016} and then we multiply this value to 80 nm to obtain an estimation of the flux in the 200-280 nm band.

\begin{figure}
    \centering
    \includegraphics[width =0.9\columnwidth]{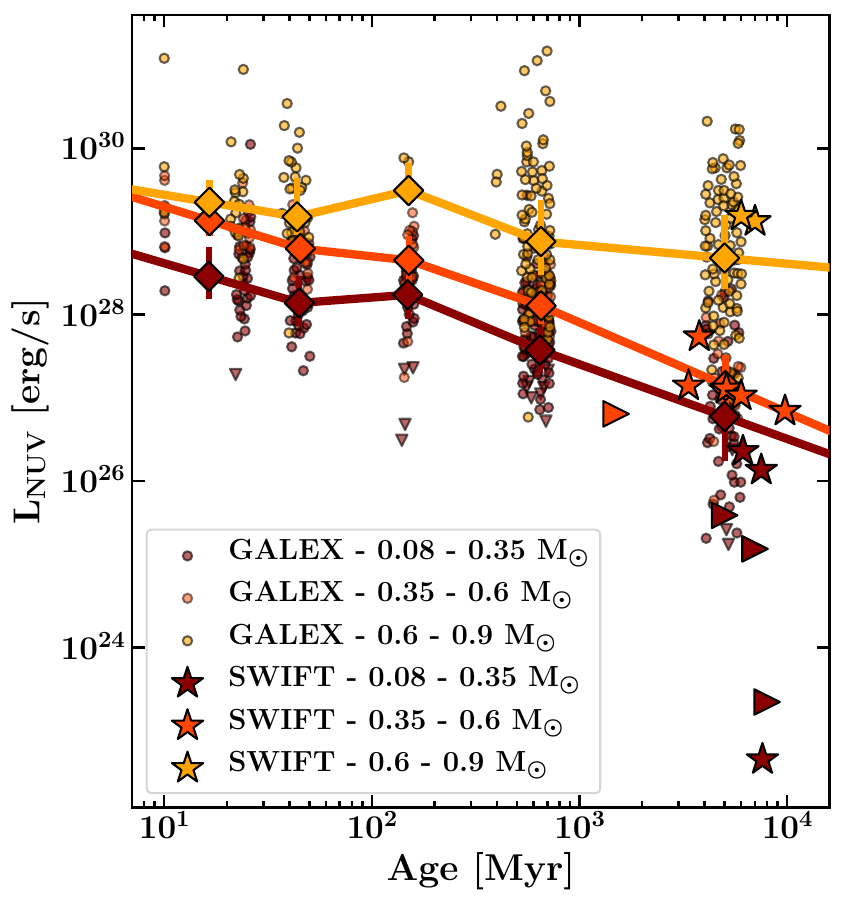}
	\caption{Dots (downward triangles) represent NUV luminosities estimated using GALEX data (upper limits) as reported in R23 for late-M-dwarfs (red), early-M-dwarfs (orange) and K stars (yellow). Diamond symbols represent the median NUV luminosity values at each age bin (same color coding) and solid lines are piece-wise linear interpolations among consecutive time bins. Star symbols represent the NUV luminosities of our selected sample stars, as measured from Swift/UVOT observations, at the time corresponding to the age of these stars reported in the literature. The same color coding highlights late-M-dwarfs (red), early-M-dwarfs (orange) and K stars (yellow). Stars with only a lower limit estimate of their age are shown with red and orange rightward triangles.}
 \label{evolution}
\end{figure}

We show in Fig.~\ref{evolution} GALEX NUV luminosities, computed from the corresponding fluxes obtained trough the method described above, as a function of the stellar age. The red, orange and yellow symbols correspond to late-M-dwarfs, early-M-dwarfs and K stars, respectively. The dots are GALEX measurements, while diamonds represent the median luminosity for each age bin. The NUV evolutionary tracks, represented by solid lines in Fig.~\ref{evolution}, are obtained by linear interpolation in between the median values. In this work, we assume that a star in our sample follows a track parallel to the evolutionary tracks of its stellar mass range normalized at the NUV luminosity measured from Swift observations at the current age (star symbols in Fig.~\ref{evolution}). Therefore, we obtain for each star the evolution of its NUV luminosity which is used to estimate, at each epoch, the inner and the outer boundaries of the UHZ.  

To address the uncertainties in stellar ages, for each target we assumed three evolutionary tracks anchored to the measured NUV luminosity. These three evolutionary tracks correspond to the estimated age and the upper and the lower ages determined by the errors in Table \ref{tab:sample}. For host stars observed by Swift which have only a lower limit estimate of their age (triangle symbol in Fig.~\ref{evolution}), we consider two NUV evolutionary tracks, namely we normalize the NUV evolution by adopting the lower limit of their age or the age of the Milky Way (11 Gyrs). These two assumptions determine two evolutionary tracks with the former having a smaller NUV luminosity than the latter at comparable ages.

\begin{table}
\centering
	\begin{tabular}{cccccccc
    	}
    	\hline
    	\hline
    	                           %
        Star & Spectral &$M_{\star}$ & $T_{\rm eff}$ & $d$ & Age \\ 
        & Type & $M_{\odot}$ & [K] & [pc] & [Gyr] \\


        \cline{1-7}
        Trappist-1 & M8.0 & 0.09 & 2550  & 12.43 & 7.6 $\pm$ 2.2 \\
        Teegarden &  M7.0 & 0.09&2790 &   3.83& >8\\
        GJ 1061  &  M5.5 &  0.13 &2905& 3.67 & >7 \\
        Proxima Cen  & 	M5.5 & 0.13&3050  & 1.30 & 7.5 $\pm$ 0.5\\
        LHS 1140   &  M4.5 &  0.18 &3166& 14.99  & >5 \\
        GJ 273  & M3.5 & 0.29&3382  & 3.79 &  6.12 $\pm$ 3.17\\
        GJ 163 &  M3.5 &0.40 &3399 &15.13 & 6  $\pm$ 4.0& \\
        K2-18 & M2.5 &  0.44 &3464 &  38.03 &  3.35  $\pm$ 0.45 \\
        GJ 357  & 	M2.5 &  0.35&3490& 9.44 & 9.76 $\pm$ 3.31\\
        TOI-700  &  M2.5 & 0.41&3494 & 31.13 & >1.5 \\
        GJ 832 &M2.0 & 0.44&3601& 4.96 &  5.2 $\pm$ 2.13  \\
        GJ 229A &  M0.0 & 0.54 & 3912 &5.76 & 3.77 $\pm$ 0.51 \\
        Kepler-62 &  K2.0 & 0.73&4842& 300.87 & 7.0 $\pm$ 4.0\\
        HD 40307 &  K2.0  & 0.79 &4867& 12.94& 6.0 $\pm$ 4.1 \\

    	\hline
    	\hline
	\end{tabular}
	\caption{Planetary systems hosting CHZ planets considered in this work, with the spectral type and $T_{\rm eff}$ of the host star, the distance from the Earth, and the age of the host stars. For the age we use the following references: \citet{Burgasser2017} (Trappist-1), \citet{Zech2019} (Teegarden's star), \citet{West2008} (GJ 1061), \citet{Beech2017} (Proxima Centauri), \citet{Dittmann} (LHS 1140),  \citet{Gaidos2023} (GJ273, GJ 832, GJ 357, GJ 229A), \citet{Tuomi2013} (GJ 163), \citet{Engle2023} (K2-18),  \citet{Gilbert2020} (TOI-700), \citet{Borucki2013} (Kepler-62), \citet{Delareza2023} (HD 40307). }
\label{tab:sample}
\end{table}

\section{Results}
\label{label:results}
\begin{figure*} 
\centering
\includegraphics[width=0.215\textwidth]{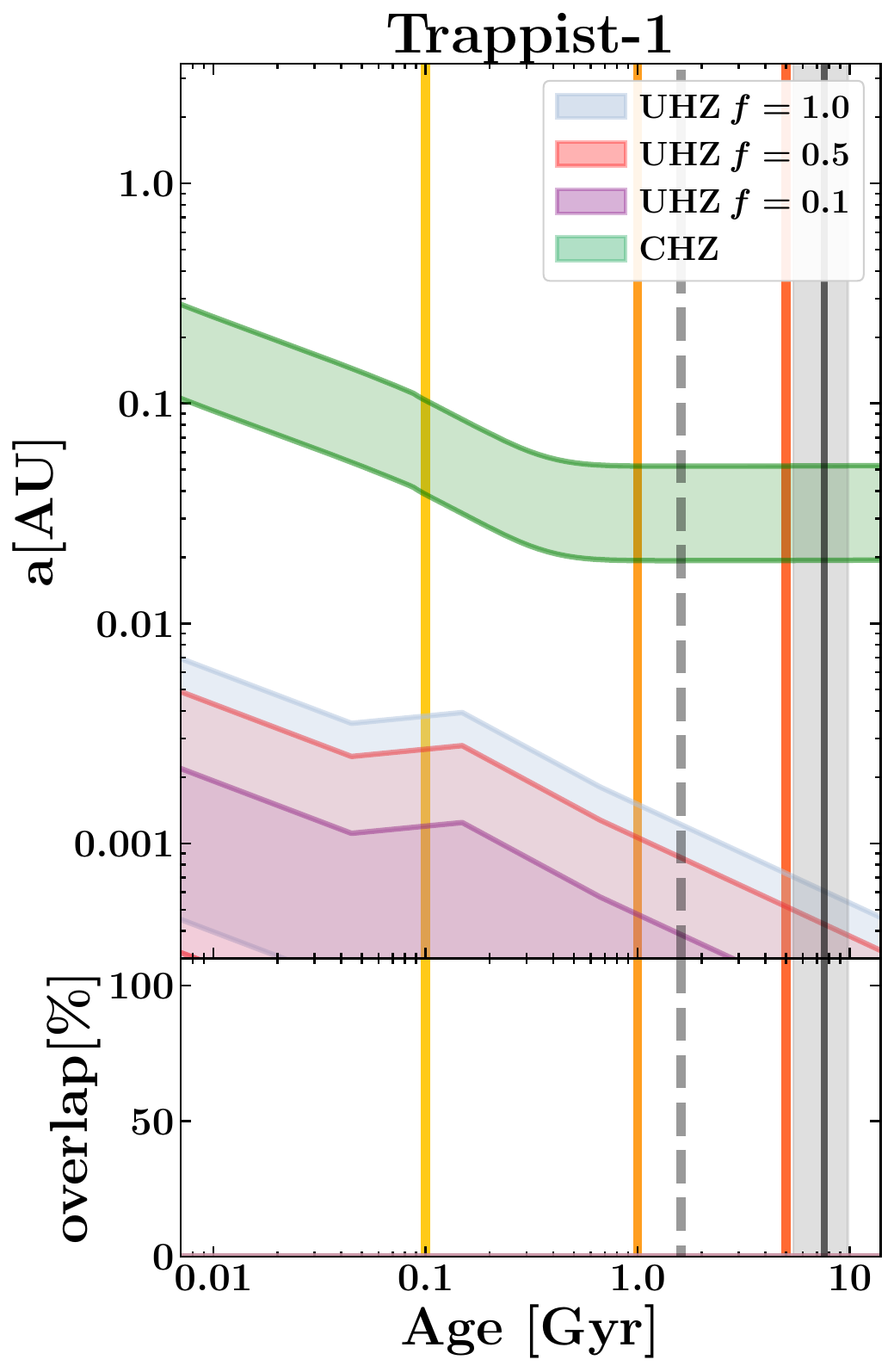}\hspace{-0.2em}
\includegraphics[width=0.18\textwidth]{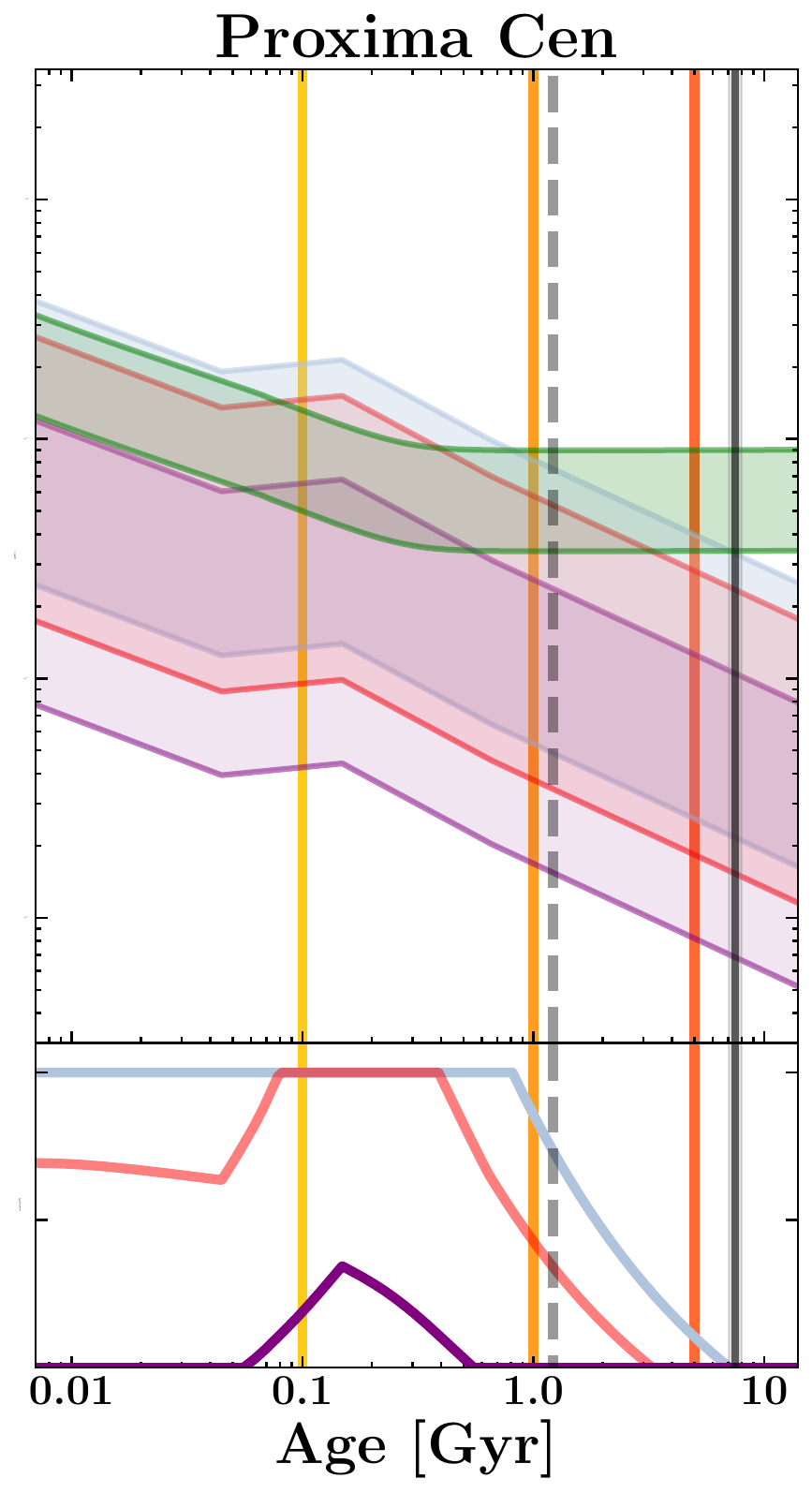}\hspace{-0.2em}
\includegraphics[width=0.18\textwidth]{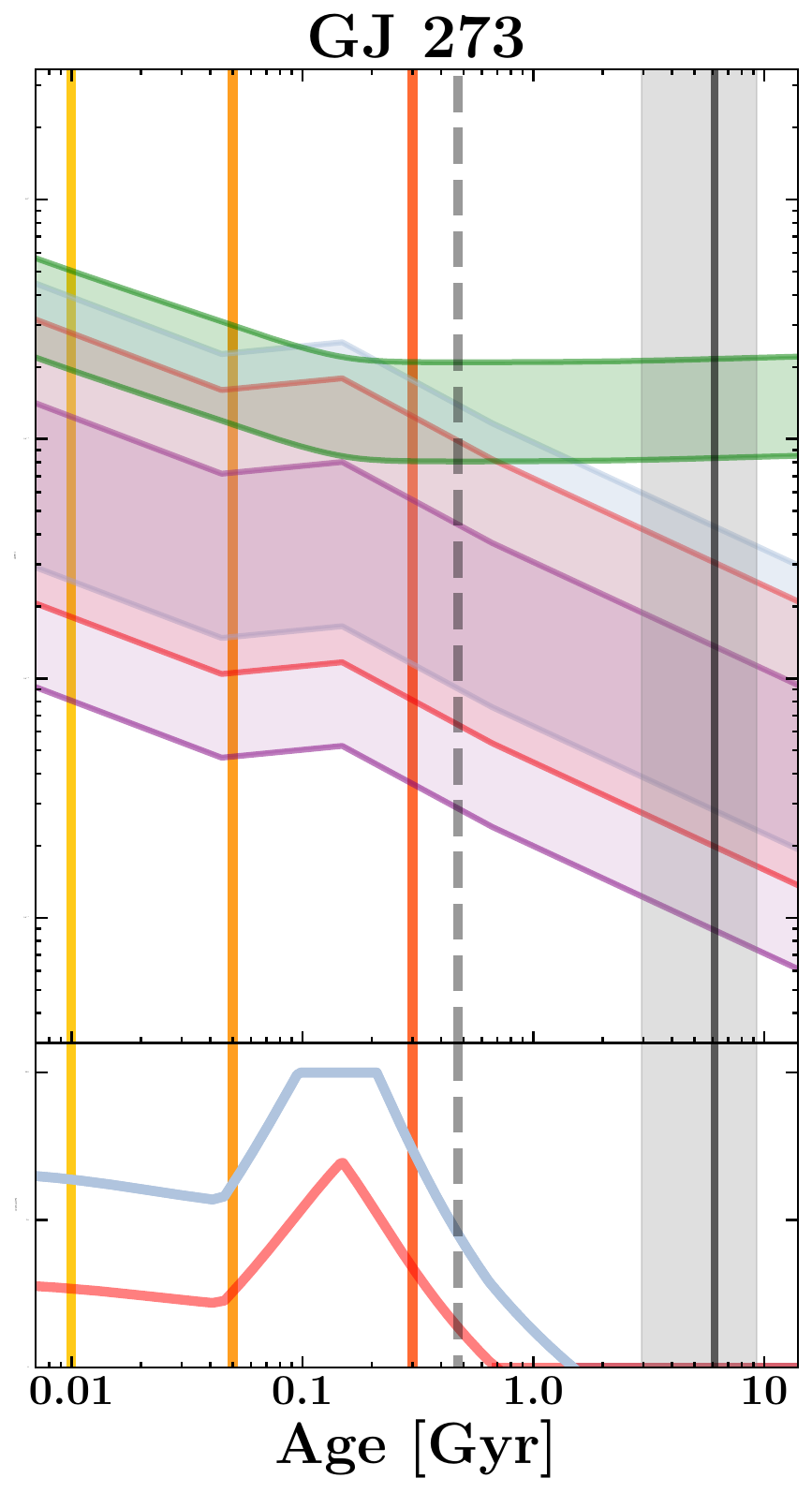} \hspace{-0.4em}
\includegraphics[width=0.18\textwidth]{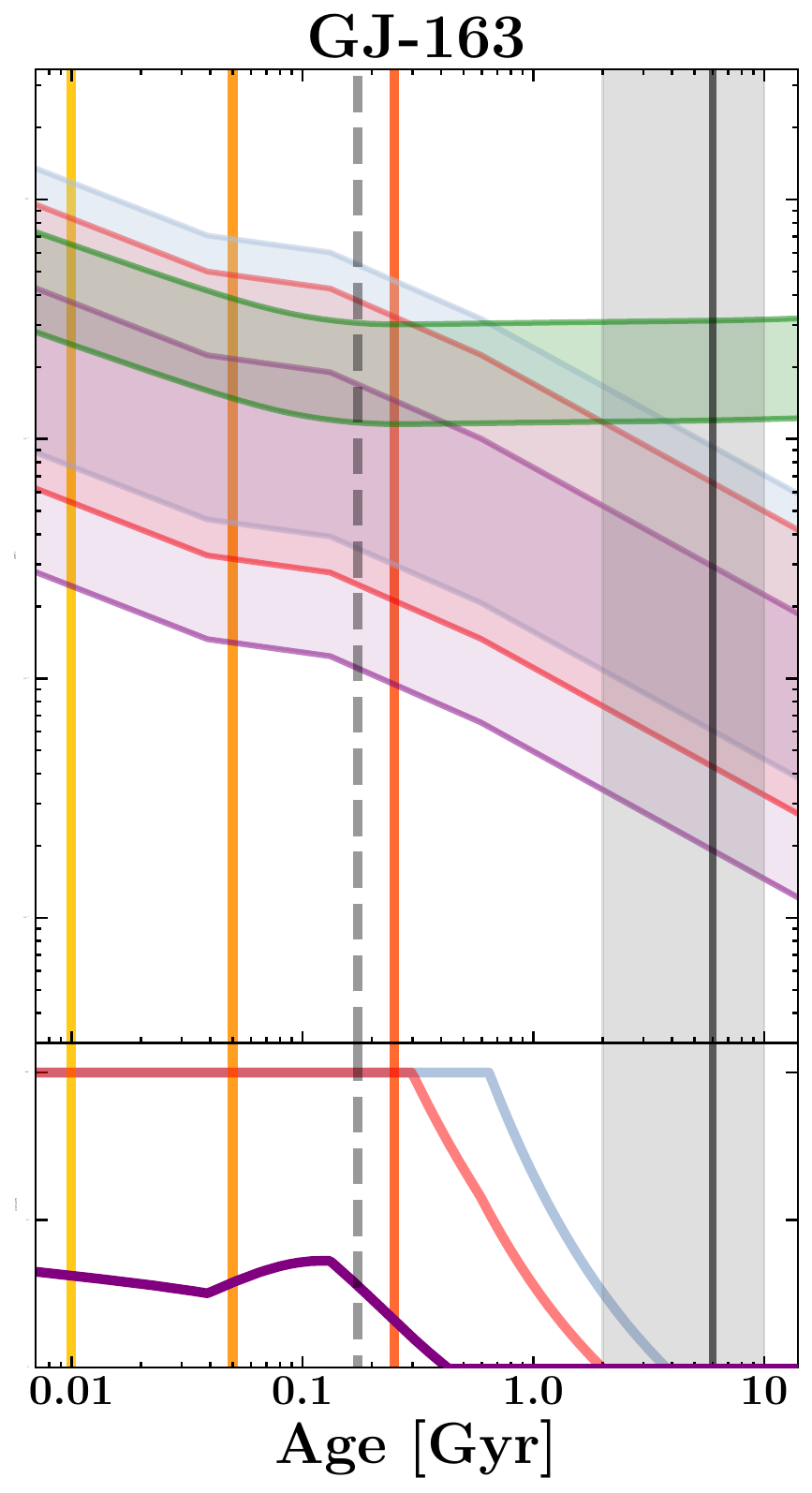} \hspace{-0.4em}
\includegraphics[width=0.18\textwidth]{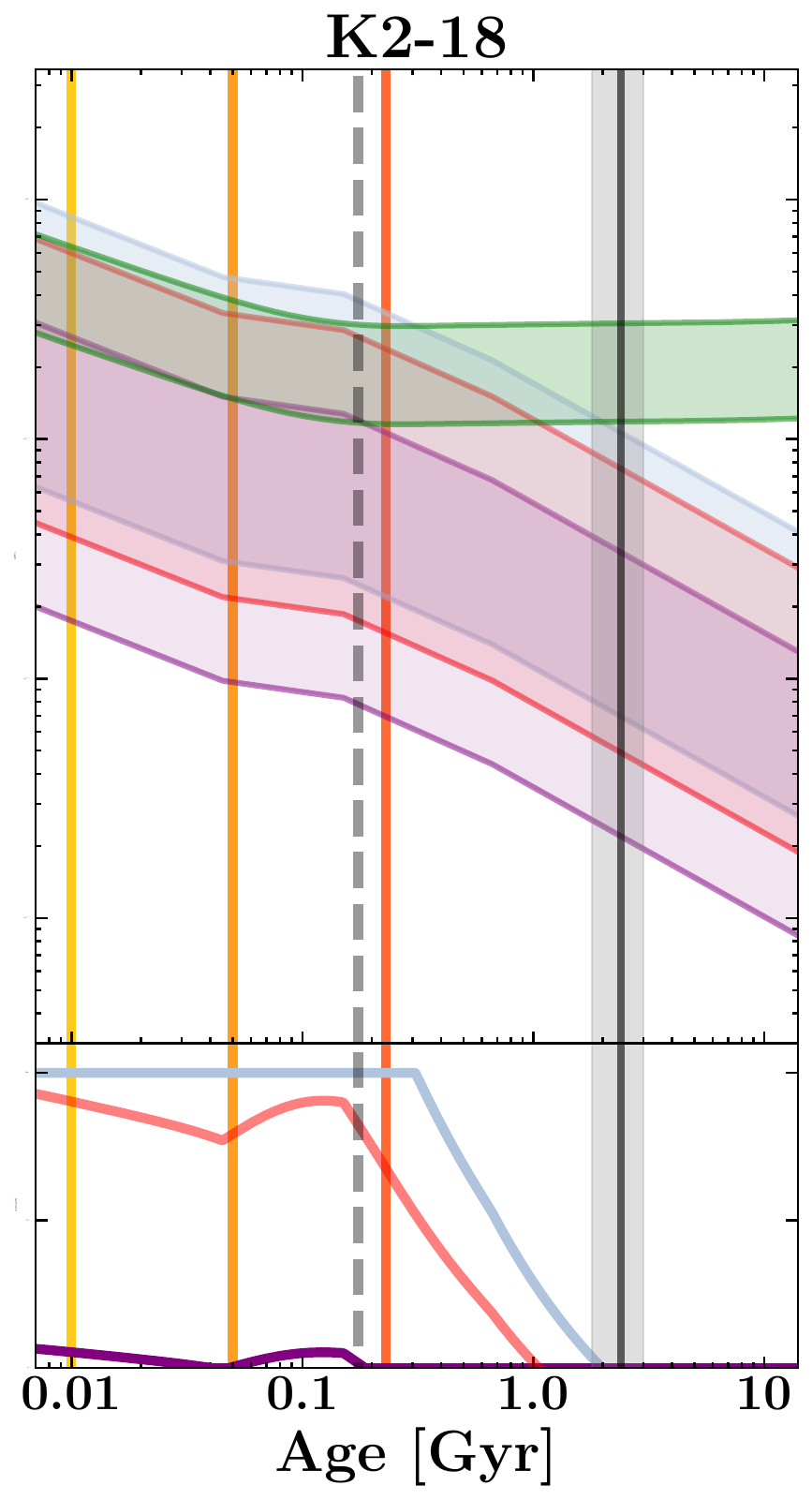} \\
\includegraphics[width=0.215\textwidth]{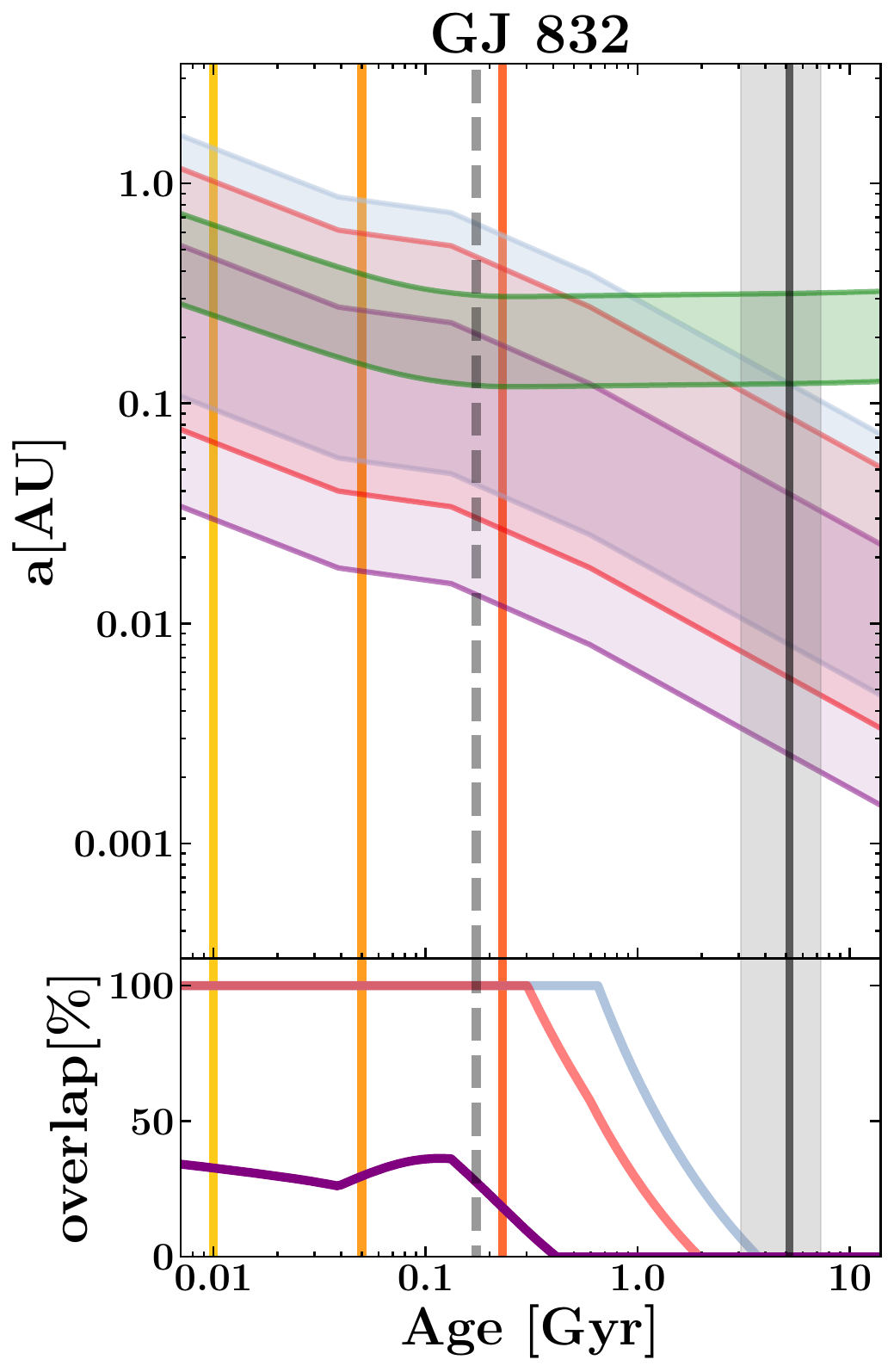}  \hspace{-0.4em}
\includegraphics[width=0.18\textwidth]{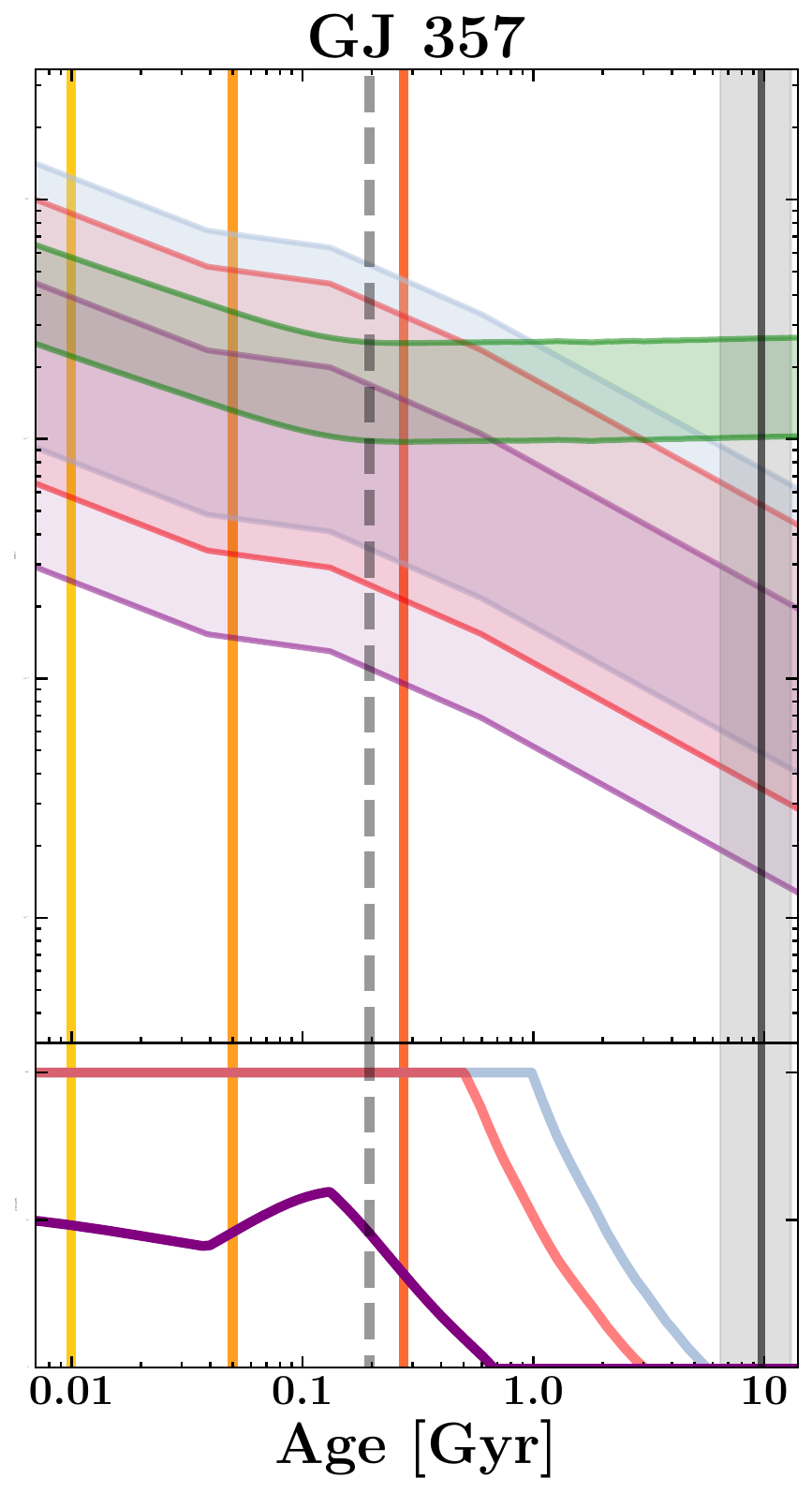} \hspace{-0.4em}
\includegraphics[width=0.18\textwidth]{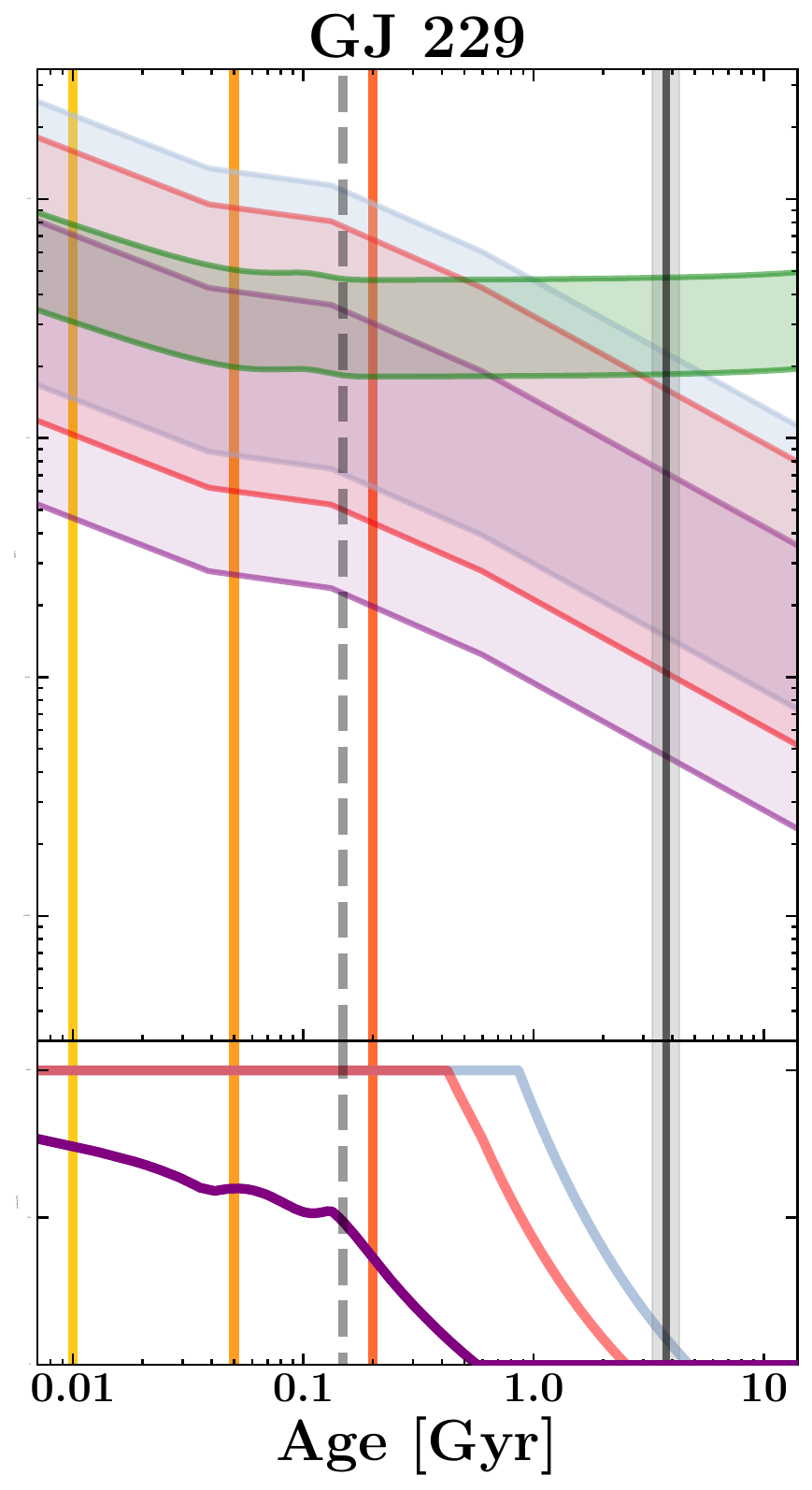} \hspace{-0.4em}
\includegraphics[width=0.18\textwidth]{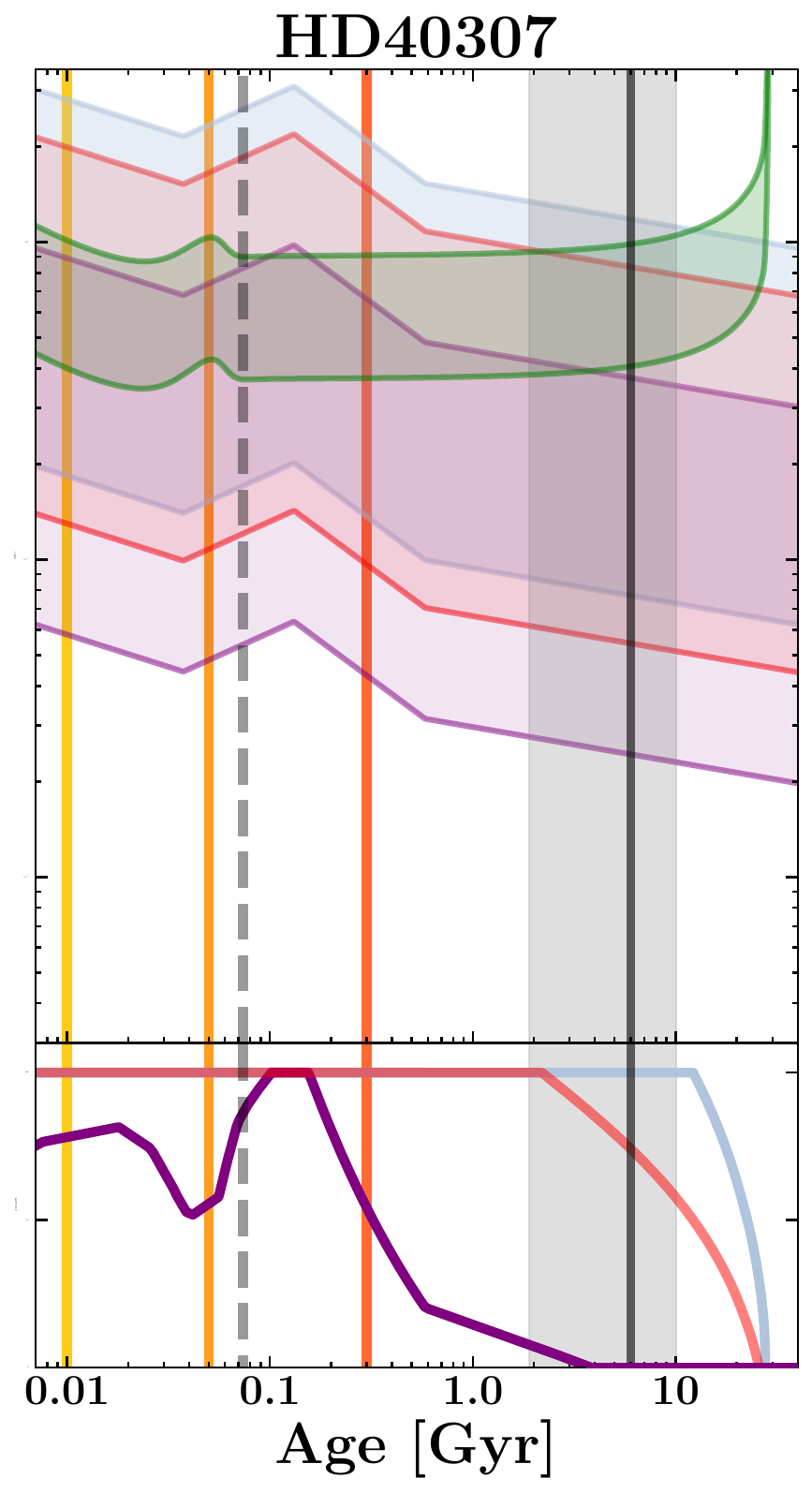}\hspace{-0.4em}
\includegraphics[width=0.18\textwidth]{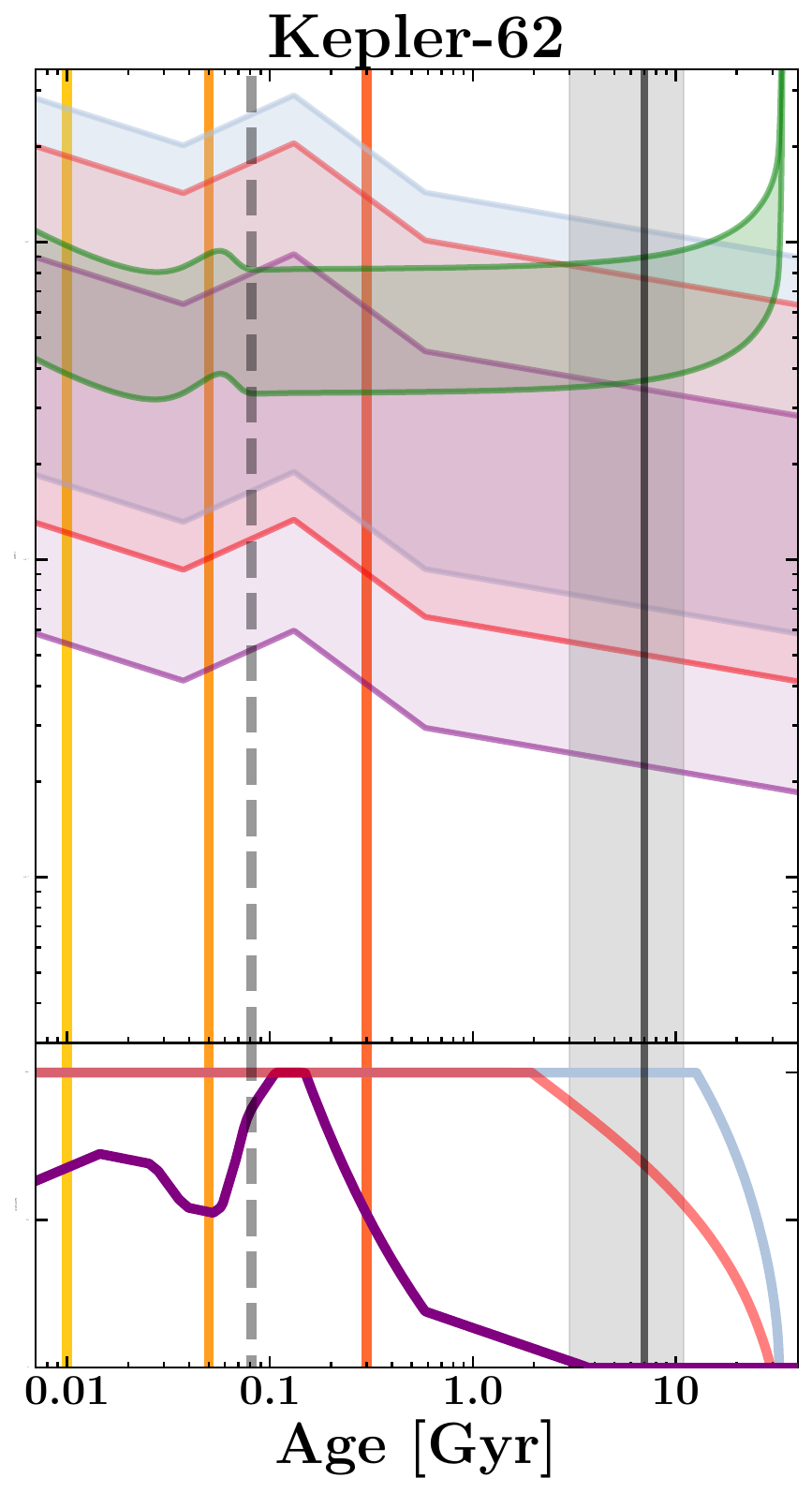} 
    \caption{Top panels: evolution of CHZ and UVH for Trappist-1, Proxima Centauri, GJ 273, GJ 163, K2-18, GJ 832, GJ 357, GJ 229A, HD 40307 and Kepler-62. In each plot the vertical gray line represents the host star age found in the literature and the relative errors (the gray shaded region around the gray line). Yellow, orange and dark orange lines represent the
ages at which fraction of systems with at least one rocky
planet without a primordial envelope in the CHZ is equal to 20\%, 30\%, 40\% for each
system (according to NGPPS model). The dashed line represents the time where each star settled onto the main sequence. Bottom panels: evolution of the fraction of the radial extension of the CHZ that is overlapping the radial extension of the UVZ for three different values of the NUV atmospheric transmission. }
    \label{fig:super}
\end{figure*}

The top panels in each plot in Fig.~\ref{fig:super} show the evolution of CHZ (green shaded area) and UHZ (violet, red, grey shaded areas) for the host stars in our sample with estimated age. Following S23, the UHZ is defined for three different NUV atmospheric transmission values: 100\% (grey), 50\% (red) and 10\% (violet). These free-parameter values correspond to atmospheres with a pressure and a composition that allow 100\%, 50\% and 10\% of the host star's NUV radiation to reach the planetary surface. The transmittivity of an atmosphere in the NUV band depends primarly on the atmospheric concentration of NUV absorbers. A transmittivity of 100\% indicates an absence of NUV absorbers. Assuming the Archean Earth's atmosphere model from \citet{Cnossen2007}, a transmission of 50\% (10\%) approximately corresponds to CO$_2$ surface concentrations of $10^{18}$ ($10^{19}$)\,cm$^{-3}$.

In each plot, the vertical gray line represents the host star age estimated in the literature. The bottom panels represent the fraction of overlap between the CHZ and the UHZ regions as a function of time for three different values of the NUV atmospheric transmission. 

\begin{figure*}
    \centering
    \includegraphics[width=0.63\textwidth]{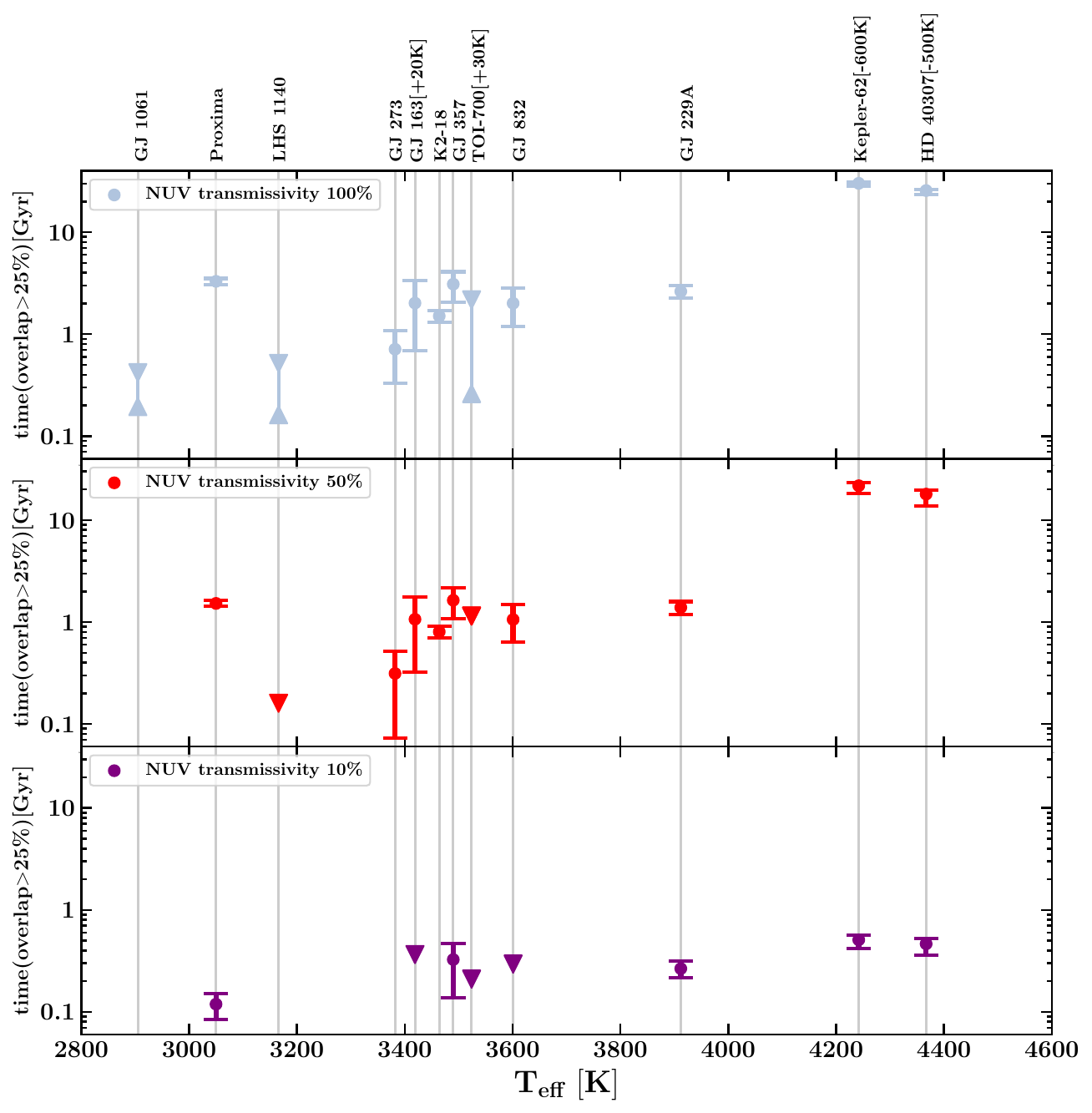}
    \caption{Time during which for each host star we have a fractional overlap greater than 25\% (violet, red, grey dots) between the radial extension of CHZ and UVH for the three different atmospheric NUV transmission (10\%, 50\%, 100\% respectively). Triangle symbols represent the upper and lower limits.}
    \label{fig:25percento}
\end{figure*}

Fig.~\ref{fig:25percento} shows, for each star, the amount of time during which the radial extension of the CHZ and UHZ overlap $>25$\%  as a function of the host effective temperature. Violet, red, and grey dots correspond to the three different atmospheric NUV transmissions, namely 10\%, 50\%, 100\% respectively. As explained in Sect. \ref{sec:chz}, we assumed three evolutionary NUV tracks for stars with an estimated age with an error and with these tracks we estimated respectively the dots and the corresponding errors in Fig.~\ref{fig:25percento}. Triangles in Fig.~\ref{fig:25percento} represent hosts with only a lower limit estimate of their ages. As explained in Sect. \ref{sec:chz},
we assumed for these cases two evolutionary NUV tracks by normalizing to the lower limit of their ages or to the age of the Milky Way (11  Gyrs).

Note that not all planetary systems have the three color symbols because it may result that, for some values of the assumed atmospheric NUV transmission, their CHZ and UHZ never overlap. 

\section{Discussion}
\label{sec:discussion}
As already found in S23 and here shown in Fig.~\ref{fig:super}, in cold stars (T$_{\rm eff}<$3900 K) in our sample, such as Trappist-1, GJ 273, Proxima Centauri, K2-18, GJ 832 and GJ 357, the present--day CHZ and the UHZ are not overlapping (vertical lines in Fig.~\ref{fig:super}). 

For GJ 229A (3912 K) today there is a relatively small (10\%) overlap between the radial extension of CHZ and UHZ assuming a 100\% of atmospheric NUV transmission, while for two K-stars in our sample (Kepler-62 and HD 40307) their CHZ is almost completely superimposed to their UHZ (for 100\% and 50\% transmission). As a consequence, the planets lying in the CHZ of these two systems should have an atmosphere sufficiently transparent to NUV radiation in order to trigger, at the present epoch, the formation of RNA precursors according to the chemistry proposed by \citet{Rimmer2018}.

The level of intersection between CHZ and UHZ (for any assumed value of the atmospheric transmission) was higher in the past for most systems and particularly the extent of the intersection was larger for hotter stars in our sample. An intersection between CHZ and UHZ (assuming a fully to half transparent atmosphere) may have existed around all the stars in our sample during the first $\simeq 1-2$\ Gyrs of their evolution. The only exceptions are the coldest (T$_{\rm eff} \lesssim 2800$ K) M-dwarfs such as Trappist-1 and Teegarden. For GJ 229A, GJ 357, Kepler-62 and HD 40307b (T$_{\rm eff} \gtrsim$ 3500 K) there was also a relatively long-lasting period in the past when they experienced a non-negligible intersection between their CHZ and UHZ with the most conservative assumption on the atmospheric transmission - violet regions. 

For stars with an effective T$_{\rm eff} \lesssim 3400$ K, such as GJ 273, GJ 1601 and LHS 1140, more than 25\% of radial extension of CHZ overlaps the grey (red) UHZ for a maximum of $\sim$700 Myr (300 Myr). An exception is represented by Proxima Centauri (T$_{\rm eff}$=3050 K), for which more than 25\% of the CHZ overlaps with the grey (red/violet) UHZ for $\sim$3.3 Gyr ($\sim$1.5/0.12 Gyr). Given the controversial debate regarding the age of Proxima Centauri (also concerning the true origin of the triple system $\alpha$ Cen, \citealt{Feng2018}), we explored if its unique level of UHZ-CHZ intersection with respect to other M-dwarfs change with the age value reported by \citet{Bazot2016} (4.8 Gyr). This age estimation leads to a reduction of the level of intersection, but remains significantly higher than in other similar M-dwarfs (more than 25\% of the CHZ overlaps with the grey (red/violet) UHZ for $\sim$2.1 Gyr ($\sim$1.0/0.0 Gyr)).

Around stars with  $3500 K \lesssim$ T$_{\rm eff}  \lesssim 4000$ K in the past there were epochs during which $>$25\% of the radial extension of the CHZ was superimposed even to the radial extension of their UHZ assuming the most conservative 10\% atmospheric transmission. For example, for GJ 357 more than 25\% of the CHZ overlaps with the grey/red/violet UHZ for $\sim$3.0/1.6/0.3 Gyr. For K star in our sample (Kepler-62 and HD 40307), the more than 25\% of the CHZ is superimposed to the violet UHZ until $\sim$500 Myr after the star settled on the main sequence (see panel Kepler-62 and HD 40307 in Fig.~\ref{fig:super}), but then the level of intersection is less than 25\% and at $\sim$3-4 Gyr completely disappears. For these stars, the intersection between CHZ and the grey/red UHZ persists until today and continues in the future until the post-main sequence phase of the stars (20-30 Gyrs). 

It is worth providing some mean estimations regarding the time of overlap between CHZ and UHZ for different types of stars. First of all, we consider late-M-dwarfs (excluding stars with effective temperature less than 2800 K, i.e. Trappist-1 and Teegarden's star): GJ 1061, LHS 1140, GJ 273 and Proxima Centauri. The average duration in which the radial extension of CHZ and UHZ overlap by more than 25\% is equal to $\sim$1.3/0.5/0.03 Gyr if the atmospheres have 100/50/10\% transmission in the NUV band\footnote{In this calculation we assume the upper limit of the overlap for GJ 1061 and LHS 1140. Assuming the lower limit, the average duration in which the radial extension of CHZ and UHZ overlap by more than 25\% is equal to $\sim$1.1/0.4/0.03 Gyr if the atmospheres have 100/50/10\% transmission in the NUV band.}.

The same calculations for early-M-dwarfs (GJ 163, K2-18, GJ 357, TOI-700, GJ 832 and GJ 229A) and K star (Kepler-62 and HD 40307) give respectively the values of $\sim$2.2/1.2/0.2 Gyr\footnote{In this calculation we assume the upper limits for TOI-700 and for GJ 163 and GJ 832 for the case of 10\% transmission in the NUV band. The same calculation assuming the lower limits provides the mean value of 1.9/1.0/0.1 Gyr} and 28/20/0.5 Gyr if the atmospheres has 100/50/10\% transmission in the NUV band.

We notice that in the present work we neglected any transformation of the planetary atmosphere during the stellar evolution possibly induced by the absorbed UV radiation and/or by potential biota or other effects.  In principle, a sudden change (on a geological time-scale) in the transmission of the atmosphere would allow a planet to "switch" between lines of different colors (bottom panels in Fig.~\ref{fig:super}) during the evolution of the system. 

Though a detailed discussion of these issues is beyond the scope of this work, in Fig.~\ref{fig:super} we reported a set of lines (yellow, orange and dark orange) that provides an estimated time for each target to host a rocky planet without a thick primordial atmosphere (composed of H/He) with a given probability. To obtain these lines, we exploited The New Generation Planetary Population Synthesis (NGPPS, \citealt{Burn2021}) simulations for low mass stars (from 0.1 to 0.7$M_{\odot}$). This planetary formation model includes planetary migration, N-body interactions between embryos, accretion of planetesimals and gas, and the long-term contraction and loss of the gaseous atmospheres. Each simulation provides at each time-step of the planetary evolution from 1 Myr to 10 Gyr many planetary parameters (such as orbital distance, planetary mass, envelope mass, planetary radius) for all the simulated planetary systems. For each time-step we calculated the fraction of systems with at least one rocky planet without a primordial envelope in the CHZ. In Fig.~\ref{fig:super} we report with yellow, orange and dark orange lines the age at which this fraction is equal to 20\%, 30\%,  40\% for each system. Moreover we add a dashed line that represents the time where each star settled onto the main sequence.

For a detailed discussion about the assumptions regarding NUV-triggered abiogenesis and the flux threshold used to define the UHZ, we refer to S23. There, we concluded that all exoplanets discovered within the classical CHZ around M-dwarf stars are not currently receiving NUV flux at levels sufficient to support the prebiotic chemistry as described in \citet{Rimmer2018}. Moreover, earlier studies \citep{Rimmer2018, Gunther2020, 2017AsBio..17..169R} have suggested that the vast majority of cold stars are, in all respects, too faint (or not sufficiently flaring) in the NUV band to trigger abiogenesis via cyanosulfidic chemistry. Generalizing this argument, we are led to conclude that M--dwarfs, which constitute the most abundant stellar population in the Galaxy \citep[$\simeq 75$\%,][]{Bochanski2010}, may not emit sufficient NUV radiation to initiate the onset of life, at least according to the specific prebiotic pathway considered here. 
In the present paper we arrived at a different conclusion, though. We assert that, when examining the evolution of NUV luminosity in M--dwarfs, most of these cool stars are indeed capable of emitting an appropriate amount of NUV photons during the first 1-2 billion years of their lifetimes. Our results suggest that the conditions for the onset of life according to the specific prebiotic pathway we consider may be or may have been common in the Galaxy. Indeed, in the present work we demonstrated that an intersection between CHZ and UHZ could exist (or could have existed) around all stars of our sample at different stages of their life, with the exceptions of the 
coolest M--dwarfs ($T\lesssim$ 2800\,K, notably Trappist-1 and Teegarden’s star). Our conclusions are further supported by Swift NUV observations of the young M--dwarf AU MIC \citep{Chavali2022}.

Recently, \citet{Rimmer2023} proposed the possibility of using exoplanets as laboratories to test the NUV-triggered origin of life scenario. This is possible because, in the next decades, exoplanetary research will be able to carry out statistical studies on the presence of biosignatures in the atmosphere of exoplanets. This distribution could indicate the most plausible environments to form life and thus the most plausible mechanism for the origin of life. 
In this work, we demonstrated that this intriguing scenario will have to consider the evolution of the environments where planetary systems evolve. 
For example, if we will find life around not-too-cold M--dwarfs (T$_{\rm eff}$> 2800 K) we can not reject the NUV-triggered abiogenesis, since it is possible that it occurred on their CHZ planets (if the atmosphere was not too opaque to NUV radiation) during the early stages of stellar evolution when these stars can be sufficiently brighter in the NUV band. 

\begin{figure}
    \centering\includegraphics[width=0.39\textwidth]{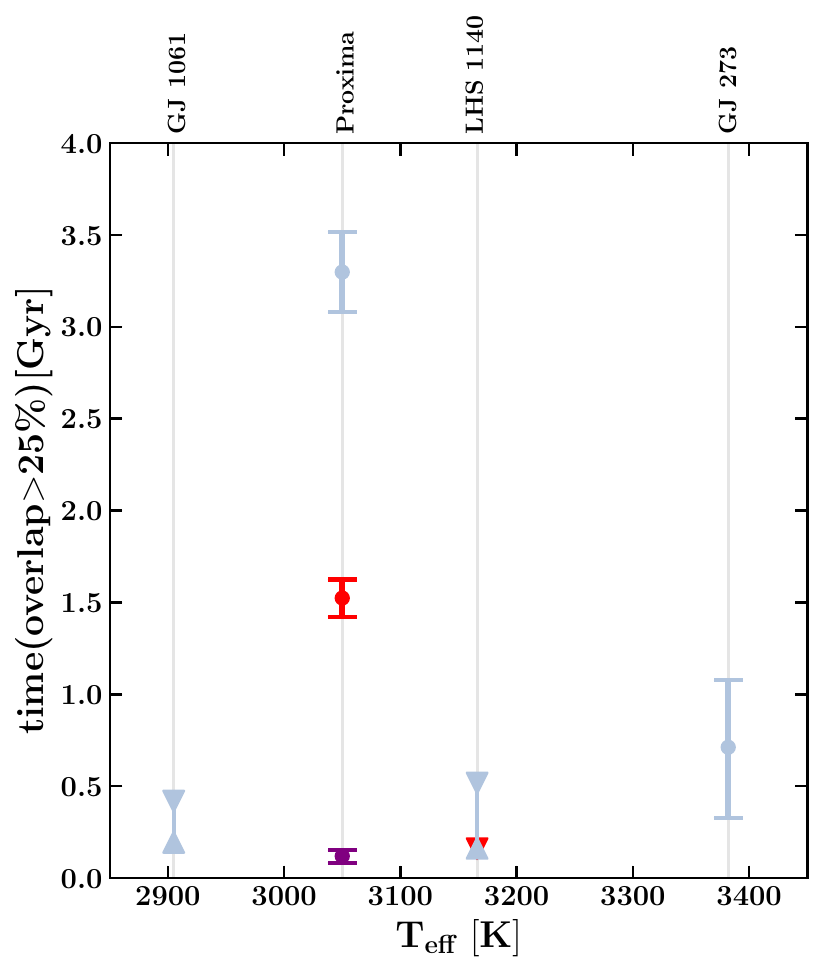}
    \caption{Time during which for each host star with effective temperature <3400 K we have a fractional overlap greater than 25\% (violet, red, grey dots) between the radial extension of CHZ and UVH for the three different atmospheric NUV transmission (10\%, 50\%, 100\% respectively). Triangle symbols represent the upper limit and lower limit for the stars for which we have a lower limit estimation of the age. For Proxima Centauri, compared to stars in our sample with similar effective temperature ($\lesssim$3400 K) such as GJ 273, LHS 1140 and GJ 1061, UHZ and CHZ had a more time-extended overlap. In particular, more than 25\% of the CHZ overlaps with the grey (red/violet) UHZ for $\sim$ 3.3 Gyr ($\sim$ 1.5 Gyr/120Myr).  }
    \label{fig:25percento-sub}
\end{figure}

To conclude we emphasize the case of Proxima Centauri b, the closest exoplanet to Earth orbiting inside the CHZ. While currently NUV flux is too low to trigger abiogenesis, UHZ and CHZ had a strong overlap (see Fig.~\ref{fig:25percento-sub}) in the past, highlighting the possibility that conditions to form life - through cyanosulfidic chemistry - have been present.

\section*{Acnowledgements}
  This study made use of data supplied by the UK {\it Swift} Science Data Centre at the University of Leicester. Part of this work is based on archival data, software, or online services provided by the Space Science Data Center-ASI. RS acknowledges Sergio Campana and Claudia Scarlata for the suggestions regarding the Swift-UVOT data analysis and Giuseppina Micela for the discussion about the evolution of the UHZ. RS acknowledges the support of the ARIEL ASI/INAF agreement n. 2021-5-HH.0 and the support of grant n. 2022J7ZFRA - Exo-planetary Cloudy Atmospheres and Stellar High energy (Exo-CASH) funded by MUR - PRIN 2022. FR acknowledges the support from the Next Generation EU funds within the National Recovery and Resilience Plan (PNRR), Mission 4 - Education and Research, Component 2 - From Research to Business (M4C2), Investment Line 3.1 - Strengthening and creation of Research Infrastructures, Project IR0000012 – "CTA+ - Cherenkov Telescope Array Plus". This publication makes use of The Data \& Analysis Center for Exoplanets (DACE), which is a facility based at the University of Geneva (CH) dedicated to extrasolar planets data visualisation, exchange and analysis. DACE is a platform of the Swiss National Centre of Competence in Research (NCCR) PlanetS, federating the Swiss expertise in Exoplanet research. The DACE platform is available at https://dace.unige.ch.
\section*{Data Availability}
The data underlying this article are available in the Swift Data Archive Mirror, see: {\url{https://www.ssdc.asi.it/mmia/index.php?mission=swiftmastr}}.








\bsp	
\label{lastpage}
\end{document}